\documentclass[a4paper,11pt]{article}

\pdfoutput=1
\usepackage{jheppub}
\usepackage{array}
\usepackage{multirow}
\usepackage{amsmath}
\usepackage{makecell}
\usepackage{enumerate}

\definecolor{green}{rgb}{0.1,0.8,0.2}

\usepackage{cancel}
\usepackage{amsmath,amssymb} 
\usepackage{graphicx} 
\usepackage[dvipsnames,x11names]{xcolor} 

\usepackage{hyperref}
\usepackage{cleveref}
\usepackage{float}
\usepackage{subcaption}

\makeatletter
\newcommand{\footnoteref}[1]{\protected@xdef\@thefnmark{\ref{#1}}\@footnotemark}
\makeatother

\begin{document}

	\preprint{}
	\title{A classical Bousso bound for higher derivative corrections to general relativity}
	\author[a]{Sayantani Bhattacharyya}
\affiliation[a]{School of Physical Sciences, National Institute of Science Education and Research, An OCC of
Homi Bhabha National Institute, Jatni-752050, India}

	\author[b]{, Parthajit Biswas}
    \affiliation[b]{Department of Physics, Ramakrishna Mission Vivekananda Educational and Research Institute, Belur Math, Howrah 711202, India}

   \author[c]{, and Nilay Kundu}
   \affiliation[c]{Department of Physics, Indian Institute of Technology Kanpur, Kalyanpur, Kanpur 208016, India}
	\emailAdd{sayanta@niser.ac.in, parthajitbiswas8@gmail.com, nilayhep@iitk.ac.in}

	\abstract{ Focussing on theories for which the higher derivative terms are considered as small corrections in the Lagrangian to Einstein's two-derivative theory of general relativity (GR), we prove the classical version of the covariant entropy bound (also known as the Bousso bound) in arbitrary diffeomorphism invariant gravitational theories. Even if the higher derivative corrections are treated perturbatively, we provide instances of specific configurations for which they can potentially violate the Bousso bound. To tackle this obstruction, we propose a modification in the Bousso bound that incorporates the offending contributions from the higher derivative corrections.  We argue that the modified Bousso bound that we propose holds to all orders in the higher curvature corrections. Our proposed modifications are equivalent to replacing the Bekenstein-Hawking area term by Wald's definition (with dynamical corrections as suggested by Wall) for the black hole entropy. Hence, the modifications are physically well motivated by results from the laws of black hole mechanics in higher derivative theories.
}


\maketitle

\section{Introduction} \label{sec:intro}

The holographic principle \cite{tHooft:1993dmi, Susskind:1994vu} is based on a physical expectation that the number of degrees of freedom in a closed region should be bounded from above by some geometric properties (like the area) of the boundary of the region. Rigorously formulating this physical expectation in terms of some mathematical inequality is a formidable task. Even before the holographic principle was conceptualized, the first proposal of such a bound was made by the seminal work of Bekenstein in \cite{Bekenstein:1980jp} to rescue the second law of thermodynamics from inconsistencies in the presence of black holes. For the validity of this bound, certain conditions were specified beyond which the bound would fail, e.g., in strongly gravitating situations of a collapse of a thermodynamic system. Moreover, in curved space-time, by general diffeomorphism invariance, the concept of the boundary of a system or radius measuring its size becomes confusing. 

A more precise and covariant version of this bound, called the Bousso bound or the covariant entropy bound, was conjectured in \cite{Bousso:1999xy} \footnote{This was influenced by a similar bound proposed in \cite{Fischler:1998st} in the context of cosmology, and was also motivated by holography, see \cite{Bousso:2002ju} for a discussion on this.}. This bound applies to null surfaces (defined as the light-sheet) generated by non-expanding light rays coming out orthogonally from a co-dimension two space-like surface. It is a statement involving an inequality-type relation between the geometry (i.e., the area of the co-dimension two space-like surfaces) and the matter entropy in its neighbourhood. In \cite{Flanagan:1999jp, Bousso:2003kb}, a stronger version of this bound, known as the generalized covariant entropy bound, was proved in Einstein's two-derivative theory of gravity 
coupled with arbitrary (but presumably minimally coupled) matter field. Here, the light sheet under consideration was restricted between two space-like surfaces, one in the past of the other. One should note that these covariant versions of the Bousso bound \cite{Bousso:1999xy, Flanagan:1999jp, Bousso:2003kb}, including the original Bekenstein bound \cite{Bekenstein:1980jp}, are all valid in the classical domain. However, in \cite{Casini_2008}, a version of the Bekenstein bound was constructed to incorporate quantum effects using properties of relative entropy in relativistic quantum field theories. Similarly, in \cite{Strominger:2003br, Bousso_2014, Bousso:2014uxa} \footnote{In \cite{Franken:2023ugu}, quantum Bousso bound has been extended to JT gravity.}, a rigorous formulation of the entropy bounds \cite{Bousso:1999xy, Flanagan:1999jp, Bousso:2003kb} was given based on quantum information theorems involving von Neumann entropy of semi-classical matter. 


Now, all such progress discussed above was restricted to two-derivative general relativity. However, general relativity is not a complete description of nature; it is an effective theory at best, describing phenomena at low energy gravitational scales. It is believed that any UV complete theory of gravity, in its low energy limit, will typically add higher derivative corrections (possibly an infinite number of them) to the gravity Lagrangian. Drawing inspiration from its original motive, the Bousso bound should be valid even in gravitational theories beyond general relativity if it is indeed a universal principle of nature. In this note, our primary goal is to examine whether this expectation is plausible. In other words, we would like to verify if the intricacies of the higher derivative corrections to general relativity can be manipulated in a way such that an appropriate, and hence a well-motivated modification to the Bousso bound can be constructed, which will make it valid in such broader class of gravitational theories. We will be modest enough to restrict ourselves to classical theories of gravity with minimally coupled matter satisfying sensible energy conditions as used in \cite{Bousso:1999xy, Flanagan:1999jp, Bousso:2003kb}. 

To answer this question, we first need to know how Einstein's equation enters all the previous proofs in two derivative theories of gravity. The line of argument starts from some bound on the entropy of a matter by its total energy content (expressed in terms of the classical matter stress tensor). Next, the stress tensor in the matter sector is related to the geometry of the space via Einstein's equation. So, by appropriate use of Einstein's equation, it is possible to convert the inequality between the entropy and the energy in the matter sector to an inequality involving the entropy and the geometry of the surrounding space of a null light sheet of any space-like hypersurface.

In most cases, while proving a classical version of the covariant entropy bound (i.e., constraining the entropy of the matter by its energy content), the starting point is itself an assumption \cite{Flanagan:1999jp, Bousso:2003kb}. In other words, the Bousso bound applies to those matter states where the entropy and the energy satisfy some particular inequalities. There are pieces of evidences that all reasonable matter would satisfy such bounds \cite{Flanagan:1999jp, Bousso:2003kb} \footnote{In \cite{Bousso_2014}, this starting inequality (in some form, not directly relatable to the classical ones as presented in \cite{Flanagan:1999jp, Bousso:2003kb}) could be proved using theorems of information theory. But, in this note, we shall restrict ourselves to the proofs of classical Bousso bound, where the starting relation between the entropy and energy of the matter is just another reasonable assumption and probably cannot be proved with our present field theoretic technology applicable to higher derivative theories of gravity.}. In our work, we shall assume that all these inequalities between the entropy of matter and its energy are valid even in the presence of higher derivative corrections. The intuition behind this simple extension is that all these matter sector inequalities are local and, therefore, should hold in any background geometry, irrespective of whether the geometry satisfies any particular equation. We shall not attempt to give any further justification for our starting assumptions other than what is already presented in \cite{Flanagan:1999jp, Bousso:2003kb}.  

As mentioned before, the crucial argument in all the proofs of the Bousso bound lies in the relation between the energy of the matter sector and the curvature of the space-time. This is precisely the relation that changes because of the higher derivative corrections due to the use of the equations of motion. Here, we shall concentrate only on the modifications (if any) needed to account for these higher derivative terms in the equations of motion.

In this context, it is worth mentioning the recent work in \cite{Matsuda_2021}, which are the precursors to our work \footnote{Also see \cite{Zhang:2022yvv, Zhu:2023xbn} for specific generalization of \cite{Matsuda_2021}.}. In \cite{Matsuda_2021}, the authors have addressed the same question we are trying to ask here. Still, their analysis is restricted to some particular cases of higher derivative theories and sometimes in a setup of spherical symmetry. In all these cases, the bound had to be modified by promoting the area term in the gravitational side of the inequality to an expression the same as that of Wald entropy with JKM \footnote{In dynamical situations, Wald entropy suffers from certain ambiguities, known in the literature as JKM \cite{Jacobson:1993vj, Jacobson:1995uq} ambiguities. For Gauss-Bonnet theory, there exists an entropy known as Jacobson-Myers(JM) \cite{Jacobson:1993xs} entropy, which does not suffer from such ambiguities. The expression of JM entropy appears in place of area for the proof of covariant entropy bound for Gauss-Bonnet theory\cite{Matsuda_2021}. For arbitrary higher curvature theories of gravity, JKM ambiguities can be fixed \cite{Wall:2015raa, Bhattacharya:2019qal, Bhattacharyya:2021jhr, Biswas:2022grc, Hollands:2022fkn, Davies:2023qaa, Bhattacharjee:2015yaa} by treating the dynamics as linearized perturbation around stationary black hole solutions, and the resulting entropy function is known as Iyer-Wald-Wall (IWW) entropy. IWW entropy will appear in place of the area in our proof of covariant entropy bound for arbitrary higher curvature theories of gravity.} ambiguities fixed. In the rest of the note, whenever we refer to Wald entropy, we always mean Wald entropy with JKM ambiguities fixed.  

Such a modification is well anticipated from the Black hole thermodynamics or generalized second law in two-derivative and higher derivative theories of gravity. Black holes are like thermal states of gravity. The `area' of the event horizon (in some sense the null boundary of the casually connected space-time) plays the role of entropy in Einstein's theory, and Wald entropy replaces it if the theory has higher derivative corrections \cite{Bardeen:1973gs, Bekenstein:1973ur,  Bekenstein:1974ax, Wald:1993nt, Iyer:1994ys, Wall:2009wm, Wall:2011hj, Sarkar:2013swa,Kolekar_2012,Bhattacharjee:2015qaa}\footnote{For a complete list of references see the recent reviews \cite{Wall:2018ydq,Sarkar_2019}}. Suppose we loosely interpret the `Bousso type bound' as a statement saying that gravitational entropy puts an upper bound on the total entropy content in a region. Then, it is natural to guess that in the presence of higher derivative corrections, the area term in the gravitational side of the Bousso-inequality has to be replaced by the expression of Wald entropy. Also, though it is a different context, the algebraic manipulation used to prove any `Bousso-type' bound in Einstein's gravity is very similar to the steps needed to confirm the Second Law in Black hole thermodynamics. Indeed, these algebraic steps need a detailed analysis of different terms (particularly their signs and relative order of magnitudes) in the equations of motion. Presumably, for this reason, the analysis in \cite{Matsuda_2021, Zhang:2022yvv, Zhu:2023xbn} had to restrict to specific higher derivative corrections or extra symmetries (like spherical symmetry in \cite{Matsuda_2021}).
 
Though our final answer about the appropriate modification of Bousso bound is the same as that of \cite{Matsuda_2021, Zhang:2022yvv, Zhu:2023xbn}, our analysis differs from theirs in two crucial aspects. Firstly, our proof works for arbitrary higher derivative corrections if they are diffeomorphically invariant. Secondly, we have assumed that the higher derivative corrections are always small compared to the scale of curvature generated by the two derivative theory itself and, therefore, could always be treated perturbatively\footnote{In contrast, for the particular cases of $f(R)$ gravity and Gauss-Bonnet theory, the authors of \cite{Matsuda_2021} have been able to argue that modified Bousso bound is valid for finite higher derivative coupling. Though it is undoubtedly an exciting and important mathematical statement about Gauss-Bonnet theory and $f(R)$ gravity,  the higher derivative couplings are certainly small at the scale of our present-day experiments, and this is the scale where our starting assumption about the bound between entropy and energy in the pure matter sector is most likely to be true.}.
 
Naively, one might wonder if any perturbative correction to the underlying equations can affect the inequality at all. After appropriate rearrangement, all inequalities can be cast as a statement about the sign of some expression. By definition, perturbative corrections to the underlying equations cannot change the sign if the expression has a finite non-zero value at the leading order. In other words, in a typical scenario, higher derivative corrections are too small to violate the original Bousso bound without any modification. Higher derivative corrections become significant only if we can construct a geometry where the Bousso bound of two derivative theories is almost saturated. In this note, we have been able to identify possible geometric configurations where higher derivative corrections can potentially violate the Bousso bound. Once we have this configuration, it immediately suggests the modification needed in the Bousso bound to take care of the offending terms. As a check of consistency, once we evaluate these modifying terms, in the particular case of \cite{Matsuda_2021}, it perfectly agrees with their result.

In this context, we would also like to mention the analysis in \cite{Bousso:2015mna}, where the authors have proposed a very general conjecture (quantum focussing conjecture or QFC) valid for any quantum matter system coupled to gravity. Though the bulk of their analysis is done in a theory of Einstein's gravity, they have some comments about how it could be generalized to higher derivative theories, at least for situations where the geometry around the light-sheet is that of a Killing horizon with linearized perturbation. The authors have further shown that in an appropriate `hydrodynamic' limit, their conjecture reduces to the covariant Bousso bound in two derivative theories. 
However, we have not chosen this route to make our point here. We did not start from QFC in higher derivative theory, and we are always in a hydrodynamic limit where, in the matter sector, the concept of entropy density makes sense. Treating the higher curvature terms in effective field theory ($\alpha^2$ expansion) and assuming a light sheet which is `a small perturbation (denoted as $\epsilon$) around a Killing horizon', we have proved Bousso bound which is valid up to arbitrary order in $\epsilon$ as well as in $\alpha^2$ expansion.


In the rest of this introduction, we shall describe the setup of our calculations along with a more quantitative summary of our final result.

 \subsection{Setup and the final result}\label{subsec:setup}
 We consider a $D$ dimensional space-time with a metric $g_{\mu\nu}$ satisfying Einstein's equation with higher derivative corrections and the dominant energy condition. Let $L$ be the null hypersurface generated by null geodesic congruence starting from a connected $(D-2)$ dimensional spatial surface $B$ and terminated on another connected $(D-2)$ dimensional spatial surface $B^\prime$. We choose the affine parameter $v$ on each of the null geodesics so that $v=0$ on $B$ and $v=1$ on $B^\prime$. Let $(x^1,x^2,...,x^{d-2})$ be the spatial coordinates on $B$, we get a natural coordinate system $(v,x^1,x^2,...,x^{d-2})$ on $L$ by simply imposing the condition that $(x^1,x^2,...,x^{d-2})$ are constants for each of the null geodesic generator. The null vector on $L$ is $k^\mu\equiv(\partial/\partial v)^\mu$. $L$ is said to be a \textit{light-sheet} when the expansion $\theta\leq 0$ everywhere on $L$. 
 
We assume that under appropriate conditions, a $D$ dimensional vector $s^\mu$ denotes the entropy flux associated with the matter sector. Working with the mostly-positive signature of the metric, we then define a local function $s(v,x)$ on the light sheet $L$ as $s(v,x)=-k_\mu s^\mu|_L$. Hence, the entropy passing through $L$ is given by (see appendix A of \cite{Flanagan:1999jp})
\begin{equation}\label{eq:defSL}
S_L=\int_B d^{D-2}x \sqrt{h(x)} \int_0^1 d v\, s(v,x)\,{\cal A}(v,x) \, ,
\end{equation}
where, $h(x)$ is the determinant of the induced metric on $B$, and ${\mathcal A}(v,x)$ is defined as follows
\begin{equation}\label{eq:Avdef}
{\cal A}(v,x)\equiv \exp\left[\int_0^v d\lambda\,\theta(\lambda,x)\right] \, ,
\end{equation}
such that $\theta=\nabla_\mu k^\mu$ is the expansion of the null geodesic congruence.

The covariant entropy bound or the Bousso bound is a precise bound on $S_L$ determined by the space-time geometry. In $D$ dimensional general relativity, under appropriate assumptions, this can be precisely written as 
\begin{equation}\label{eq:BBoundGR}
S_L  \leq \frac{1}{4\hbar G_N}\Big[A(B)-A(B^\prime)\Big]\,, 
\end{equation}
where $A(B)$ is the area of the $(D-2)$ dimensional space-like slices $B$, and the same for $B'$ \footnote{The bound in \eqref{eq:BBoundGR} is slightly different (actually stronger) than the original one in \cite{Bousso:1999xy}. In \cite{Bousso:1999xy}, this bound was originally stated for a light sheet associated with the initial slice $B$, but not truncated at a finite $v$-distance away at $B'$. Therefore, $v$ could run until $v=-\infty$ unless a singularity is encountered. In \cite{Flanagan:1999jp}, the authors extended that proof applicable to a stronger version as given in \eqref{eq:BBoundGR}, which we will be interested in. In \cite{Strominger:2003br} and \cite{Bousso:2003kb} \eqref{eq:BBoundGR} has been termed as the generalized covariant entropy bound, or the generalized Bousso bound.}. In this note, we aim to generalize this Bousso bound for arbitrary higher curvature theories of gravity treated as small corrections to general relativity. Our main result is to propose a modified Bousso bound for such theories beyond GR under an appropriately modified set of assumptions. We will show that \eqref{eq:BBoundGR} can be generalized to the following
\begin{equation}\label{eq:finalresult}
S_L  \leq \frac{1}{4\hbar G_N}\Big[S_{IWW}(B)-S_{IWW}(B^\prime)\Big]\,, 
\end{equation}
where, $S_{IWW}$ is denoted as the Iyer-Wald-Wall entropy. In the context of dynamical black holes in higher derivative theories of gravity, following the seminal work of \cite{Wald:1993nt, Iyer:1994ys}, recently, in \cite{Wall:2015raa, Bhattacharyya:2021jhr}, it has been shown that $S_{IWW}$ satisfies the linearized second law perturbatively around a Killing horizon. From these works, it is known that $S_{IWW}$ is defined completely in terms of the geometry of the space-time around the null horizon (with slightly broken Killing symmetry) of the dynamical black hole.

When discussing the Bousso bound associated with a light sheet of space-like slices, there is no reason for one to consider laws of black hole mechanics. However, it is intriguing to note the similarity of the modifications needed in the two apparently different principles. Previously, we have already mentioned an intuitive reason for this; i.e., the RHS of the inequality in \eqref{eq:finalresult} resembles gravitational entropy. However, a more technical reason for this can also be offered. As shown in \cite{Bhattacharyya:2021jhr}, a crucial ingredient in establishing the linearized second law was to obtain an off-shell structure of the null projected equations of motion, i.e., $E_{\mu\nu} k^\mu k^\nu$, on the horizon. In this process, an algorithm to compute $S_{IWW}$ from the off-shell $E_{\mu\nu} k^\mu k^\nu$ was also derived for arbitrary higher derivative theories of gravity. Our goal in this paper is to identify situations where the higher derivative gravity contributions will be dominating enough to violate the Bousso bound. Furthermore, we will then use the same off-shell structure of $E_{\mu\nu} k^\mu k^\nu$ in the neighbourhood of the light sheet, now thought of as a co-dimension one null hypersurface with a slightly broken Killing symmetry. Hence, similar to the story of black hole mechanics, we get $A(B)$ in \eqref{eq:BBoundGR} substituted by $S_{IWW}$ in \eqref{eq:finalresult} as we go beyond general relativity. 

Historically, a set of assumptions has played a rather significant role in constructing a proof of Bousso bound existing in the literature. We will follow the assumptions outlined in \cite{Strominger:2003br} applicable to two derivative general relativity. The first of these assumptions is a local inequality putting an upper bound on the rate of change of matter entropy density limited by the energy flux of the matter (through its stress-energy tensor) written in appropriately chosen units as 
\begin{equation}
s^\prime(v,x)\leq \frac{2\pi}{\hbar}{\cal T}(v,x),\quad \text{where} \quad {\cal T}\equiv T_{\mu\nu}k^\mu k^\nu\,, 
\end{equation}
where a prime denotes derivative with respect to the affinely parametrized coordinate $v$. Along with this one, one also needs another mild assumption about the entropy on the initial $v=0$ slice $B$, which essentially demands that to start with, the Bousso bound is satisfied in a region infinitesimally close to $B$ at the beginning of the light sheet. We shall explain them in more detail in the relevant section. Since the statement of the Bousso bound is going to be modified to incorporate the effects of higher derivative corrections for gravitational theories beyond GR, this second assumption is also going to naturally get modified accordingly. 


As mentioned above, the higher derivative corrections we consider in this paper are arbitrary because we do not work with any particular form of the Lagrangian apart from being diffeomorphism invariant. However, we treat them as small compared to the leading Einstein-Hilbert term in an effective field theory (EFT) setup. Therefore, the gravity Lagrangian for theories under consideration can be written as 
\begin{equation}\label{gravlag}
\mathcal{L}_G = R + \sum_{m=1}^\infty\alpha^{2m} \, \mathcal{L}_{2m+2} = R + \alpha^2 \, \mathcal{L}_{HD} \, ,
\end{equation}
where $R$ being the Ricci scalar stands for the leading two-derivative Einstein-Hilbert term and $\mathcal{L}_{2m+2}$ denotes the higher derivative corrections containing $2m+2$ number of derivatives. The dimensionful coupling $\alpha$ in eq.\eqref{gravlag} controls the smallness of such higher derivative corrections. Note that $\mathcal{L}_{HD}$ includes higher curvature terms appearing with increasing number of powers of $\alpha^2$. The gravity part of the equation of motion coming from the Lagrangian \eqref{gravlag} can be written as
\begin{equation}\label{eq:graveom}
E_{\mu\nu}\equiv R_{\mu\nu}-\frac{1}{2}R g_{\mu\nu}+\sum_{m=1}^\infty\alpha^{2m} \, E^{(2m+2)}_{\mu\nu}  = R_{\mu\nu}-\frac{1}{2}R g_{\mu\nu}+ \alpha^2 \,  {\cal E}^{(HD)}_{\mu\nu} \, .
\end{equation}
We will treat the higher derivative part of the equation of motion perturbatively in $\alpha^2$ expansion. Note that the Iyer-Wald-Wall entropy $S_{IWW}$ that appears in our final result of the modified bound written in \eqref{eq:finalresult}, also, by definition, incorporates corrections to all orders in $\alpha^2$. As a consequense, in our EFT setup the modified bound will hold to all orders in $\alpha^2$ as well.


%
%
The rest of the note is organized as follows:
 In \S\ref{sec:review}, we will review Bousso bound and its proof for the two derivative theory of general relativity  - which will set the stage for further discussions. In \S\ref{sec:violation}, we will discuss the Bousso bound in the presence of higher curvature corrections; in particular, we will discuss how the Bousso bound can potentially violated in the presence of higher curvature corrections. In \S\ref{sec:modification}, we will modify the Bousso bound to account for the higher curvature terms. We will end with conclusions and future directions in \S\ref{sec:conclude}.

\section{Review of Bousso bound in Einstein's theory}\label{sec:review}
In this section, we will review the proof of the covariant Bousso bound, focussing on Einstein's theory of general relativity, as discussed in \cite{Flanagan:1999jp, Bousso:2003kb}.

Let $L$ be the light-sheet generated by a non-expanding null geodesic congruence starting from a connected $(D-2)$ dimensional spatial surface $B$ and terminated on another connected $(D-2)$ dimensional spatial surface $B^\prime$. We choose the affine parameter $v$ on each of the null geodesics so that $v=0$ on $B$ and $v=1$ on $B^\prime$. 
It turns out that we could always choose a coordinate system so that the bulk space-time near the light sheet takes the following form
\begin{equation}\label{eq:metric}
ds^2=2\, dv\, dr-r^2 X(r,v,x) dv^2+2\,r\,\omega_i(r,v,x)\, dv\, dx^i+h_{ij}(r,v,x)\, dx^i dx^j\,.
\end{equation}
Where $r=0$ is the position of the light-sheet $L$. The null generator of $L$ is $\partial_v$ with $v$ being the affine parameter (see appendix A of \cite{Bhattacharyya:2016xfs} for details). 

The derivation of Bousso bound for Einstein's theory starts from the following two assumptions. The first of these assumptions, as explained in \cite{Strominger:2003br}, is motivated by an underlying thermodynamic approximation and places a hierarchy of scales between the rate of evolution of matter entropy and its energy flux. 
\begin{equation}\label{eq:assump1} 
(i)\quad s^\prime(v,x)\leq \frac{2\pi}{\hbar}{\cal T}(v,x),\quad {\cal T}\equiv k^\mu k^\nu T_{\mu\nu}(r=0) \, ,  
\end{equation}
such that $T_{\mu\nu}$ is the matter stress tensor and prime denotes derivative with respect to $v$. The second assumption is just an assumption that the Bousso bound is satisfied locally around the initial spatial slice $B$ at $v=0$, 
\begin{equation}\label{eq:assump2} 
(ii)\quad s(0,x)\leq -\frac{\theta(0,x)}{4\hbar G_N}\,.
\end{equation}
Here, $\theta(v,x)$ is the expansion of the congruence of null geodesics on the light-sheet and in our choice of coordinates it has the following expressions
\begin{equation}\label{eq:theta}
\begin{split}
\theta(v,x) \equiv {\partial_v \sqrt{h(v,x)}\over \sqrt{h(v,x)}},~~\text{where}~~h(v,x) \equiv\text{Det}\big[h_{ij}(r=0,v,x)\big]\,.
\end{split}
\end{equation}
Using Einstein's equations and specializing in our choice of coordinates on the light sheet, ${\cal T}$ could be expressed as
\begin{equation}\label{eq:tmunu}
\begin{split}
{\cal T}=k^\mu k^\nu T_{\mu\nu}(r=0) =\frac{1}{8\pi G_N}\left( k^\mu k^\nu R_{\mu\nu} \right)_{r=0} =\frac{1}{8\pi G_N}\left(-\frac{d\theta}{dv}-\frac{1}{2}\theta^2-\sigma^{ij}\sigma_{ij}\right) \, ,
\end{split}
\end{equation}
where in the last step, we have used the Raychaudhuri equation, and $\sigma^{ij}$ is the shear of the null congruence. Substituting equation \eqref{eq:tmunu} in the first assumption \eqref{eq:assump1} we find 
\begin{equation}\label{eq:process1}
\begin{split}
s^\prime(v,x) +{1\over 4\hbar G_N}\left(\frac{d\theta}{dv}\right) \leq- {1\over 4\hbar G_N}\left[\frac{1}{2}\theta^2+\sigma^{ij}\sigma_{ij}\right] \, .
\end{split}
\end{equation}
Now integrating \eqref{eq:process1} with respect to $v$ starting from $v=0$ we get
\begin{equation}\label{eq:process2}
\begin{split}
s(v,x) +{\theta(v,x)\over 4\hbar G_N} \leq & \left(s(0,x) +{\theta(0,x)\over 4\hbar G_N} \right) - {1\over 4\hbar G_N}\int_0^{v} dv'~\left[\frac{1}{2}\theta^2(v',x)+\sigma^{ij}(v',x)\sigma_{ij}(v',x)\right] \, .
\end{split}
\end{equation}
If we further integrate the inequality \eqref{eq:process2}, first on an arbitrary constant $v$ spatial slice $\Sigma_v$ of the light-sheet (with induced metric $h_{ij}(v,x)$) and then with respect to $v$ from $v=0$ to $v=1$, the left hand side of \eqref{eq:process2} takes the following form
\begin{equation}\label{eq:process3}
\begin{split}
&\int_{v=0}^{v=1} dv\int d^{D-2}x\, \sqrt{h(v,x)}\left(s(v,x) + {\theta(v,x)\over 4\hbar G_N} \right)\\
&=\int d^{D-2}x\, \sqrt{h(0,x)}\int_{v=0}^{v=1} dv\sqrt{h(v,x)\over h(0,x)} \, s(v,x)  + {1\over 4\hbar G_N}\int d^{D-2}x \, \int_{v=0}^{v=1} dv\, \partial_v\left(\sqrt{h(v,x)}\right)\\
&= S_L +{1\over 4\hbar G_N}\int d^{D-2}x\left(\sqrt{h(1,x)} -\sqrt{h(0,x)}\right)\\
&= S_L -\left(A(B) -A(B')\over 4\hbar G_N\right) \, ,
\end{split}
\end{equation}
where we used the definition of $S_L$ from \eqref{eq:defSL}. Let us define  of ${\cal F}$ as 
\begin{equation} \label{defF}
{\cal F} \equiv S_L -\left(A(B) -A(B')\over 4\hbar G_N\right) \, . 
\end{equation}

We could apply the same set of integral operations on the RHS of the inequality \eqref{eq:process2}. Unlike the LHS, it is not possible to process these integrations further. Still, we could easily see that the final answer has to be negative since the integrand is manifestly negative definite (the first term in RHS of \eqref{eq:process2} is negative as a consequence of our assumption \eqref{eq:assump2} and the second term is just negative of full squares). So it follows that 
${\cal F}\leq0$ and hence 
\begin{equation}
S_L \leq\left(A(B) -A(B')\over 4\hbar G_N\right) \, , 
\end{equation}
which is the desired Bousso bound for Einstein's theory of two-derivative gravity.

\section{Possible violation of Bousso bound due to higher derivative corrections}\label{sec:violation}
In the previous subsection, we have seen that Einstein's equation has been used to replace the matter stress tensor in terms of geometric quantities in equations \eqref{eq:tmunu} and \eqref{eq:process1}. Let us consider higher derivative corrections to the leading two derivative general relativity with the Lagrangian given in \eqref{gravlag}. The combined equations of motion will also get additional contributions. We assume that the four derivative correction term $\alpha^2 \mathcal{L}_4$ as defined in \eqref{gravlag} is non-zero, therefore the equation of motion coming from $\alpha^2 \mathcal{L}_4$ dominate over all the other higher derivative correction terms\footnote{Here, $\mathcal{E}_{\mu\nu}^{(HD)}\equiv E_{\mu\nu}^{(4)}+\alpha^2 E_{\mu\nu}^{(6)}+\dots$ as defined in \eqref{eq:graveom}. If for some special case the four derivative correction term in the Lagrangian becomes zero, effectively the higher derivative corrections in \eqref{HDeqn} starts with $\alpha^4$. And the rest of the analysis would go through.}
\begin{equation}\label{HDeqn}
{\cal E}^{tot}_{\mu\nu} = R_{\mu\nu}-{1\over 2} g_{\mu\nu} R+ \alpha^2 \,  {\cal E}^{(HD)}_{\mu\nu} \, . 
\end{equation}
As mentioned before, we will not be working with any particular solutions to the equations of motion or restrict ourselves to any specific symmetry (like spherical symmetry). However, we will only consider such configurations for which the higher derivative contributions will be small compared to the intrinsic curvature scales associated with those configurations. Therefore, when evaluated on such configurations, terms with the coefficient $\alpha^2$ will always be small compared to the contributions from leading two-derivative theory. 

With this setup of higher derivative corrections in our mind, let us now track down the modifications that might invalidate the derivation of the Bousso bound presented in the previous section. It is easy to see that the addition of higher derivative corrections ${\cal E}^{(HD)}_{\mu\nu}$ to the equations of motion in \eqref{HDeqn} will modify \eqref{eq:tmunu}. Then, this modification will propagate to the integrand in the RHS of inequality \eqref{eq:process2}. The changes in \eqref{eq:tmunu} can be written down as follows 
\begin{equation}\label{eq:processHD1}
\begin{split}
{\cal T}=\frac{1}{8\pi G_N}k^\mu k^\nu \left[ R_{\mu\nu} + \alpha^2{\cal E}^{(HD)}_{\mu\nu}\right]_{r=0} =\frac{1}{8\pi G_N}\left[-\frac{d\theta}{dv}-\frac{1}{2}\theta^2-\sigma^{ij}\sigma_{ij} + \alpha^2{\cal E}^{(HD)}_{vv}\right] \, ,
\end{split}
\end{equation}
and finally \eqref{eq:process2} will become 
\begin{equation}\label{eq:processHD2}
\begin{split}
s(v,x) &+{\theta(v,x)\over 4\hbar G_N} \leq\left(s(0,x) +{\theta(0,x)\over 4\hbar G_N} \right) \\
&-{1\over 4\hbar G_N}\int_0^{v} dv'\, \left[\frac{1}{2}\theta^2(v',x)+\sigma^{ij}(v',x)\sigma_{ij}(v',x) -  \alpha^2{\cal E}_{vv}^{(HD)}(v',x)\right] \, .
\end{split}
\end{equation}

The proof for the Bousso bound crucially depends on the sign of the RHS in \eqref{eq:processHD2}, in particular the integrand in the second term, which has been modified because of the higher derivative terms in equations of motion. But in general, the sign of the term $ \alpha^2{\cal E}_{vv}^{(HD)}(v',x)$ is not fixed, which potentially obstructs a proof of Bousso bound. However, in our setup, the dimensionful couplings of the higher derivative terms are always small compared to the length scale set by the curvature of the geometry. Therefore, one might doubt if at all a small correction can ever change the sign of the dominant leading order terms. In other words, the $ \alpha^2{\cal E}_{vv}^{(HD)}(v',x)$ term will always be sub-dominant compared to the contribution from the two-derivative theory ($\theta^2$ and $\sigma^2$) and thus will lead us to the conclusion that the proof of Bousso bound will be unaffected. 

The conclusion we have just made about the validity of the Bousso bound is too quick. We will now point out a subtlety where the corrections due to the higher derivative terms can become significant enough to pose a threat to the proof of Bousso bound. 
We have in mind possible configurations as solutions to the equations of motion, for which the $\theta^2$ and $\sigma^2$ terms on the RHS of \eqref{eq:processHD2} happen to be very small. They are small enough to be of the same order of magnitude as that of $\alpha^2{\cal E}_{vv}^{(HD)}$. There is a possibility that the higher derivative corrections can overcome the leading two-derivative contributions with an appropriate sign, making the inequality in \eqref{eq:processHD2} invalid. Hence, there is a clear chance of violation of the Bousso bound. 

Let us now elaborate a bit more on such a possible scenario. If the light-sheet generator $k^\mu =\partial_v$ is proportional to a Killing vector of the space-time (at least in some finite neighbourhood of the light-sheet), then the vanishing of the shear $\sigma_{ij}$ and the expansion $\theta$ follow just from the Killing equations. One might naively think that this is the case where the sign of ${\cal E}_{vv}^{(HD)}$ matters and can potentially violate the Bousso bound. But it turns out that using the Killing symmetry of space-time, we could argue that ${\cal E}_{vv}^{(HD)}$ would also vanish \cite{Bhattacharyya:2021jhr}. Therefore, what we need here is a slight violation of the Killing condition such that the light-sheet generators have non-zero $\theta$ and $\sigma_{ij}$, but their values are comparable with ${\cal E}_{vv}^{(HD)}$.

Therefore, to replicate a prototype geometry of a light sheet with slightly broken Killing symmetries, let us assume that the metric around the light sheet has the following structure
\begin{equation}\label{eq:metpert}
\begin{split}
ds^2=2\, dv\, dr-r^2 X(r,v,x^i) dv^2 & +2\,r\,\omega_i(r,v,x^i)\, dv\, dx^i+h_{ij}(r,v,x^i)dx^i dx^j\, ,\\
\text{where} \quad 
X(r,v,x^i) &= \bar X(rv,x^i) + \epsilon~ \delta X (r,v,x^i) \, ,\\
\omega_i(r,v,x^i) &= \bar\omega_i(rv,x^i) + \epsilon~ \delta \omega_i (r,v,x^i) \, ,\\
h_{ij}(r,v,x^i) &= \bar h_{ij}(rv,x^i) + \epsilon~ \delta h_{ij} (r,v,x^i) \, .
\end{split}
\end{equation}
Here, $r=0$ is the position of the light-sheet spanned by the null coordinate $v$ and the $(D-2)$ spatial coordinates $x^i$, and both $r$ and $v$ are affinely parametrized. Note that we are not assuming any spherical symmetry in this ansatz for the metric. Also, $\epsilon$ is an arbitrarily small real parameter that could have either sign. If we set $\epsilon$ to zero, the metric will have a Killing vector of the form $ \xi^\mu\partial_\mu = v\partial_v - r\partial_r $, which on the light-sheet at $r=0$  becomes proportional to the null generator of the light-sheet $\partial_v$. Thus, for $\epsilon=0$, to satisfy the Killing equation, the barred metric components $\bar X, \bar\omega_i$, and $\bar h_{ij}$ depend on the coordinate $v$ only through the product $rv$. 

The parameter $\epsilon$ in \eqref{eq:metpert} measures the small departure of the light-sheet from satisfying the Killing conditions. Thus, in \eqref{eq:metpert}, these small fluctuations in the metric denoted by $\delta X, \, \delta \omega_i$, and $\delta h_{ij}$ have arbitrary dependence on the coordinates $(r, \, v, x^i)$. It is easy to see that on the light-sheet both $\theta$ and $\sigma_{ij}$ evaluate to terms of order ${\cal O}(\epsilon)$,
\begin{equation} \label{thetasigeps}
\theta = \epsilon\left({\bar h^{ij} \, \partial_v \delta h_{ij} \over 2} \right)_{r=0} + {\cal O}(\epsilon^2), \quad \text{and} \quad \sigma_{ij} = \left( \epsilon \, \partial_v \delta h_{ij}- \bar h_{ij} \, \theta \right)_{r=0} + {\cal O}(\epsilon^2) \, .
\end{equation}

If we do not specify the nature of the Higher derivative corrections to the gravity Lagrangian, we cannot compute ${\cal E}_{vv}^{(HD)}$ explicitly. But following the analysis of \cite{Bhattacharyya:2021jhr}
\footnote{The Killing vector $\xi^\mu$ is the generator of a boost transformation $v \to \lambda v, r \to r/\lambda$ with $\lambda$ being a constant. This is an exact symmetry of the metric for $\epsilon=0$, see \cite{Bhattacharyya:2021jhr}. With a small non-vanishing $\epsilon$, this symmetry gets broken slightly. Based on how different objects transform under this transformation, one can check that any object of the form $\partial_v A$ when evaluated on $r=0$ can be non-zero only when $A \sim {\cal O}(\epsilon)$. This also justifies \eqref{thetasigeps}. More details can be found in \cite{Bhattacharyya:2021jhr}. It should be noted that the analysis in \cite{Bhattacharyya:2021jhr} has been done in the context of a black hole space-time with a Killing horizon and the small non-stationary fluctuations around it. Nevertheless, it could be applied to any null hypersurface without any modification. In fact, in \cite{Bhattacharyya:2021jhr}, horizons are treated as special cases of light sheets with some particular final condition as the affine parameter along the null geodesics tends to infinity.} 
we could argue that generically ${\cal E}_{vv}^{(HD)}$ will also be at least of order ${\cal O}(\epsilon)$. Recall ${\cal E}_{vv}^{(HD)}$ is also multiplied by the higher derivative coupling constant denoted by $\alpha^2$. 

Now the integrand in the RHS of the inequality \eqref{eq:processHD2} has three terms, two of them ($\theta^2$ and $\sigma^2$) are of order ${\cal O}(\epsilon^2)$ whereas the higher derivative correction ${\cal E}_{vv}^{(HD)}$ is of order ${\cal O}(\alpha^2 \, \epsilon)$. So, the first two terms will be comparable to the third if we choose $\epsilon$ to be of the order of the higher derivative coupling $\alpha^2$. Therefore, we are looking for a geometric configuration for which 
\begin{equation}
\epsilon^2 \sim \alpha^2 \, \epsilon, \quad \text{such that} \quad \theta^2 \, \, \text{and} \, \,  \sigma^2 \sim \alpha^2{\cal E}_{vv}^{(HD)} \, .
\end{equation}
Since both the magnitude and the sign of $\epsilon$ is a matter of choice, we could, therefore, always construct a geometry where the integrand on the RHS of \eqref{eq:processHD2} has the opposite sign than what is required for the Bousso bound to be proved.

To summarize, we have argued that if the geometry around the light sheet is of the form of equation \eqref{eq:metpert} with $\epsilon$ of the order ${\cal O}(\alpha^2)$, then the higher derivative corrections to the equation of motion can potentially violate the Bousso bound. To construct a proof for Bousso bound for theories with higher derivative corrections, one needs to tackle this situation, and that will be the focus of our calculations in the next section. 


\section{Modification of Bousso bound to account for ${\cal E}_{vv}^{(HD)}$}\label{sec:modification}
In the previous section, we have identified the geometry \eqref{eq:metpert} where the Bousso bound could be violated because of the higher derivative correction. In this section, we want to verify if modifying the Bousso bound by some higher derivative correction can address the offending term. 

The key expression we need to handle is the sign of the integrand in the RHS of inequality \eqref{eq:processHD2}. For convenience, let us denote this integrand as {\cal M} such that 
\begin{equation}\label{eq:name}
\begin{split}
{\cal M} \equiv \frac{1}{2}\theta^2+\sigma^{ij}\sigma_{ij} -\alpha^2 {\cal E}^{(HD)}_{vv} \, . 
\end{split}
\end{equation}
The first two terms in ${\cal M}$ are always non-negative but are of order ${\cal O}(\epsilon^2)$ once evaluated on the geometry \eqref{eq:metpert}, but they are independent of $\alpha$. The form of the third term ${\cal E}^{(HD)}_{vv}$ depends on the details of the higher derivative corrections. Therefore, unlike the first two terms, it could only be evaluated explicitly in terms of the metric functions if one works with a specific form of the Lagrangian. However, in \cite{Bhattacharyya:2021jhr}, the authors, just using the symmetries of the background geometry, have been able to predict a general structure for the off-shell structure of ${\cal E}_{vv}^{(HD)}$ computed at $r=0$, if evaluated on a metric of the form \eqref{eq:metpert}. This is given by 
\begin{equation}\label{eq:Evv}
\begin{split}
{\cal E}_{vv}^{(HD)}&=-\partial_v{\cal Q}+{\cal O}(\epsilon^2) \, ,\\
\text{where} \quad & {\cal Q}\equiv  \frac{1}{\sqrt{h}}\partial_v\left(\sqrt{h}\,{\cal J}_{IWW}\right)+\frac{1}{\sqrt{h}}\partial_i\left(\sqrt{h}{\cal J}^i\right)  \, .
\end{split}
\end{equation}
Here ${\cal J}_{IWW}$ and ${\cal J}^i$ could be constructed out of the metric functions and their $v$, $r$ and $i$ derivatives, evaluated at $r=0$.  
Note that the order ${\cal O}(\epsilon^0)$ piece in ${\cal J}_{IWW}$ will not contribute to ${\cal Q}$ since its $v$ derivative vanishes on the horizon. Consequently, the contribution of ${\cal J}_{IWW}$ to ${\cal Q}$ starts from order ${\cal O}(\epsilon)$. On the other hand, ${\cal J}^i$ is always of order ${\cal O}(\epsilon)$. Another important feature of \eqref{eq:Evv} is that this form of ${\cal E}_{vv}^{(HD)}$ is true to all orders in the higher derivative coupling $\alpha^2$.

Now, in ${\cal M}$, the first two terms on the RHS of \eqref{eq:name} are of order ${\cal O}(\alpha^0\epsilon^2)$. So all the terms in ${\cal E}_{vv}^{(HD)}$ which are quadratic or of higher power in $\epsilon$, are always suppressed compared to the first two terms because of the factors of $\alpha^2$ multiplying higher derivative corrections. It is only the order ${\cal O}(\epsilon)$ piece in ${\cal E}_{vv}^{(HD)}$ that could be larger than the first two terms in ${\cal M}$ if $\epsilon$ is as small as $\alpha^2$.

To cure this problem, we propose a certain modification in the statement of the Bousso bound. We proceed by adding appropriate integration of $\partial_v{\cal Q}$ (the order ${\cal O}(\epsilon)$ term of ${\cal E}_{vv}^{(HD)}$ in \eqref{eq:Evv}), to both sides of the inequality \eqref{eq:processHD2} as follows,
 \begin{equation}\label{eq:processHD3}
 \begin{split}
 &s(v,x) +{\theta(v,x) + \alpha^2{\cal Q}(v,x)\over 4\hbar G_N} \leq \left(s(0,x) +{\theta(0,x) + \alpha^2{\cal Q}(0,x)\over 4\hbar G_N} \right) \\
& -{1\over 4\hbar G_N} \int_0^{v} dv'\bigg(\frac{\theta^2(v',x)}{2}+\sigma^{ij}(v',x)\sigma_{ij}(v',x) -\alpha^2 \bigg({\cal E}_{vv}^{(HD)}(v',x)+ \partial_{v'}{\cal Q} (v',x)\bigg)\bigg) \, .
 \end{split}
 \end{equation}
To derive \eqref{eq:processHD3}, one should first note the following trivial relation
\begin{equation}\label{eq:processHD3a}
 \begin{split}
 {\alpha^2\over 4\hbar G_N}\int_0^{v} dv' \, \partial_{v'}{\cal Q} (v',x) = {\alpha^2\over 4\hbar G_N}\left( {\cal Q}(v,x) - {\cal Q}(0,x) \right) \, ,
 \end{split}
 \end{equation}
and add this \eqref{eq:processHD3a} to \eqref{eq:processHD2}. After that, a simple algebraic manipulation would lead us to \eqref{eq:processHD3} \footnote{ It is important to note that ${\cal Q}(v,x)$ in \eqref{eq:processHD3} will receive contributions from all orders in $\alpha^2$ present in \eqref{gravlag}. This is because \eqref{eq:Evv} is an exact expression in $\alpha^2$. To make it more explicit, let us assume that we are working with the gravity Lagrangian in EFT expansion to ${\cal O} (\alpha^4)$, say, for example, given by $\mathcal{L}  = R + \alpha^2 R_{\mu\nu}R^{\mu\nu}+ \alpha^4 R^3 = R + \alpha^2( \, R_{\mu\nu}R^{\mu\nu} + \alpha^2 \, R^3)$. It is easy to see that ${\cal Q}(v,x)$ for that case will be given by ${\cal Q} = {\cal Q}_1 + \alpha^2 \, {\cal Q}_2$, such that ${\cal Q}_1$ and ${\cal Q}_2$ will be given by \eqref{eq:Evv}, receiving contributions from both $R_{\mu\nu}R^{\mu\nu}$ and $R^3$ terms respectively. This plays an important role in establishing that our proposed modification of the Bousso bound will hold true to all orders in $\alpha^2$.}. 

The last term within the integration on the RHS of \eqref{eq:processHD3} is always of order ${\cal O}(\alpha^2\epsilon^2)$ or smaller, i.e.,
\begin{equation}
\alpha^2\bigg({\cal E}_{vv}^{(HD)}(v',x)+ \partial_{v'}{\cal Q} (v',x)\bigg) \sim {\cal O}(\alpha^2\epsilon^2) \, .
\end{equation}
Therefore, although this term is not of any definite sign, its magnitude can never be larger than the first two positive definite terms in the integrand since both of them are of ${\cal O}(\alpha^0\epsilon^2)$
\begin{equation}\label{eq:processHD3b}
 \begin{split}
\frac{1}{2}\theta^2(v',x)+\sigma^{ij}(v',x)\sigma_{ij}(v',x) \gg  \alpha^2\bigg({\cal E}_{vv}^{(HD)}(v',x)+\partial_{v'}{\cal Q} (v',x) \bigg) \, .
 \end{split}
 \end{equation}
Hence, it follows that the second line in the RHS of \eqref{eq:processHD3}, involving the $v$-integration, is negative definite. We are using two perturbative parameters, one is $\alpha^2$ and the other one is $\epsilon$. Now, let us explain why no higher order terms coming from $\epsilon$ expansion or $\alpha^2$ expansion can change the sign of the second line in the RHS of \eqref{eq:processHD3}. We have taken care of order $\mathcal{O}(\epsilon)$ and $\mathcal{O}(\epsilon^2)$ terms coming from the $\mathcal{O}(\alpha^0)$ term of the effective Lagrangian, we don't need to worry about $\mathcal{O}(\epsilon^3)$ or higher order terms coming from the $\mathcal{O}(\alpha^0)$ term of the effective Lagrangian - that would be already suppressed. We have also taken care of $\mathcal{O}(\epsilon)$ terms coming from the $\mathcal{O}(\alpha^2)$ terms of the effective Lagrangian, these terms are of the form $\mathcal{O}(\alpha^2\epsilon)$, now in our set up $\alpha^2\sim\epsilon$, therefore these terms are of the order $\mathcal{O}(\epsilon^2)$. Order $\mathcal{O}(\epsilon)^2$ or higher epsilon order terms coming from $\mathcal{O}(\alpha^2)$ term of the effective Lagrangian would be suppressed. Similarly, all the higher order terms in $\alpha^2$ as well as in $\epsilon$ would also be suppressed.

Now, to prove the Bousso bound, we have to determine the sign of the first term on the RHS of \eqref{eq:processHD3}, i.e., $\left(s(0,x) +(\theta(0,x) + \alpha^2{\cal Q}(0,x)) / (4\hbar G_N) \right)$. Let us remind ourselves how this term was dealt with for general relativity, for which ${\cal Q} = 0$. From the analysis reviewed in \S\,\ref{sec:review}, we can see that the assumption in \eqref{eq:assump2} took care of the first term on the RHS of \eqref{eq:process2}. We learn to adapt a similar strategy for higher derivative corrections to general relativity and decide to correct the assumption \eqref{eq:assump2}. In the case of the two-derivative theory of gravity, this assumption says that the matter entropy density at the initial slice is less than the initial rate of decrease of the infinitesimal area element on $B$. Physically, we need this assumption to ensure that the bound is not violated at the initial slice at $v=0$ by some `violent' initial data. Now, as we will modify the bound, it is expected that the constraints on the initial data will also have to be modified accordingly. From the structure of the inequality \eqref{eq:processHD3}, it is easy to see that the following modification of \eqref{eq:assump2} would work here
 \begin{equation}\label{eq:assump2HD}
 \begin{split}
 s(0,x) +{\theta(0,x) + \alpha^2{\cal Q}(0,x)\over 4\hbar G_N}  \leq 0 \, .
 \end{split}
 \end{equation}
Once we impose the assumption \eqref{eq:assump2HD}, the RHS of the inequality \eqref{eq:processHD3} is always negative and therefore the following inequality follows
  \begin{equation}\label{eq:processHD4}
 \begin{split}
 s(v,x) +{\theta(v,x) + \alpha^2{\cal Q}(v,x)\over 4\hbar G_N} \leq&0 \, ,
  \end{split}
 \end{equation}
which would be the statement of a modified Bousso bound for theories with higher derivative correction to general relativity. 

It is essential to highlight that such a simple modification of assumption \eqref{eq:process2} (the inequality that leads to the Bousso bound in Einstein's gravity) has been possible only because the order ${\cal O}(\epsilon)$ term in ${\cal E}_{vv}^{(HD)}$ is a total $v$ derivative, i.e., $\partial_v {\cal Q}$ in \eqref{eq:Evv}. That is why the $v$ integration of the higher derivative correction evaluates precisely to some local geometric property (${\cal Q}(v,x)$) of the constant $v$ slice of the light-sheet and the local entropy density on the slice is bounded by local geometry of the slice even with the higher derivative corrections.  

Just like in the two-derivative theory, we shall further integrate \eqref{eq:processHD4} over $v$ and the spatial coordinates $\{x^i\}$ to get a bound on $S_L$.
As explained before, in the case of two-derivative theory, one of the key features of the Bousso bound is that the total matter entropy passing through the light-sheet could be bounded by the geometries of only the initial and the final spatial slices ($B$ and $B'$ respectively) and independent of how the light-sheet evolves at intermediate values of $v$. In fact, any modified bound on $S_L$ in higher derivative theories would be meaningful from the perspective of holography only if this geometric locality on the light sheet is preserved. In other words, if on ${\cal Q}(v,x)$ we perform the integral operations as done in equation \eqref{eq:process3}, the final answer should turn out to be the difference of some `density-type' geometric quantities integrated only on $B$ and $B'$. For this to happen, the structure of ${\cal Q}(v,x)$ in \eqref{eq:Evv}, predicted in \cite{Bhattacharyya:2021jhr}, is a must. We could see it as follows
\begin{equation}\label{eq:int}
\begin{split}
&\int d^{D-2}x \int_0^1 dv\,\sqrt{h(v,x)}\,{\cal Q}(v,x)\\
 =& \int d^{D-2}x \int_0^1 dv\,\partial_v\left[\sqrt{h(v,x)}{\cal J}_{IWW}(v,x)\right]  + \underbrace{\int d^{D-2}x \int_0^1 dv\,\partial_i\left[\sqrt{h(v,x)}{\cal J}^i(v,x)\right]}_{\text{goes to zero}}\\
 =& \int d^{D-2}x \sqrt{h(1,x)}\,{\cal J}_{IWW}(1,x)-\int d^{D-2}x \sqrt{h(0,x)}\,{\cal J}_{IWW}(0,x)\\
  =& \int_{B'} {\cal J}_{IWW}(1,x)-\int_B{\cal J}_{IWW}(0,x) \, .
\end{split}
\end{equation}
To go from the second to the third step, in the second term we have first interchanged the $v$ and $x$ integration and then used the fact that the spatial slices at any $v$, say $\Sigma_v$, are compact surfaces without any boundaries \footnote{In arriving at the last step, we also used the definition $\int_{B(v)} \equiv \int d^{D-2} \sqrt{h(v,x)}$.}. As a result, the spatial integration of a total derivative term - like $\partial_i\left(\sqrt{h(v,x)}{\cal J}^i(v,x)\right)$ - always vanishes,
\begin{equation}\label{eq:mani}
\begin{split}
&\int d^{D-2}x \int_0^1 dv\,\partial_i\left[\sqrt{h(v,x)}{\cal J}^i(v,x)\right] =~\int_0^1 dv\int_{\Sigma_v} d^{D-2}x \ \partial_i\left[\sqrt{h(v,x)}{\cal J}^i(v,x)\right] \\ 
& =0 \, ,  \quad  \text{for compact horizon slices} \, \, \Sigma_v \, .
\end{split}
\end{equation}

Here, we would also like to highlight the importance of the $\ \frac {1}{\sqrt{h}}\partial_i\left(\sqrt{h}{\cal J}^i\right)$ term in ${\cal Q}$, see in \eqref{eq:Evv}. The presence of this term in the off-shell structure of ${\cal E}_{vv}^{(HD)}$ in any diffeomorphism invariant theory of gravity was established in \cite{Bhattacharyya:2021jhr}. We can see the significance of this term here; had it not been of this particular form, the $x^i$-integration over a compact surface in \eqref{eq:mani} would never have given a vanishing result. 

Substituting equation \eqref{eq:int} in the appropriately integrated version of the inequality \eqref{eq:processHD4} we find
\begin{equation}\label{eq:processfinalHD}
\begin{split}
S_L\leq \frac{1}{4\hbar G_N}\int_B  d^{D-2}x \left[1 +\alpha^2{\cal J}_{IWW}(0,x)\right]- \frac{1}{4\hbar G_N}\int_{B'}  d^{D-2}x \left[1 +\alpha^2{\cal J}_{IWW}(1,x)\right] \, .
\end{split}
\end{equation}
Now, in the context of black hole thermodynamics in higher derivative theory, the RHS of the inequality \eqref{eq:processfinalHD} plays an important role. This turns out to be the candidate for gravitational entropy that reduces to the Wald entropy in a stationary situation and satisfies the second law of thermodynamics for dynamics of small amplitude away from equilibrium configurations \cite{Wald:1993nt, Iyer:1994ys, Wall:2015raa, Bhattacharyya:2021jhr}. Following these, let us define the following quantity 
\begin{equation} \label{defIWW}
S_{IWW}(v) =  \int  d^{D-2}x \, \sqrt{h(v,x)}~\left[1 +\alpha^2{\cal J}_{IWW}(v,x)\right] \, ,  
\end{equation} 
which is known as the Iyer-Wald-Wall entropy \footnote{The contribution from general relativity to black hole entropy is given by the area of the constant $v$ slices of the horizon, which is captured by the term $1$ on the RHS of \eqref{defIWW}. Also, ${\cal J}_{IWW}$ receives a contribution from the higher derivative corrections to general relativity. It should also be noted that in obtaining ${\cal J}_{IWW}$, the JKM ambiguities also get fixed.}. 
Finally using the above definition of $S_{IWW}$ in \eqref{eq:processfinalHD} we get our final result \eqref{eq:finalresult}
\begin{equation} \label{eq:finresult1}
S_L  \leq \frac{1}{4\hbar G_N}\Big[S_{IWW}(B)-S_{IWW}(B^\prime)\Big]\,.
\end{equation}
As we have mentioned below equation \eqref{eq:processHD3b}, the above inequality is valid up to arbitrary order in $\alpha^2$ expansion and up to arbitrary order in $\epsilon$ expansion.
Our final result in terms of $S_{IWW}$ is consistent with \cite{Matsuda_2021} where the Bousso bound was proved for Gauss-Bonnet theory with spherically symmetric configurations. This is obvious since $S_{IWW}$, once evaluated for the Gauss-Bonnet theory, reproduces the Jacobson-Myers entropy that appeared in the proof of Bousso bound as proposed in \cite{Matsuda_2021}. 

Before we conclude this section, let us also point out one further justification supporting the modified initial condition in \eqref{eq:assump2HD}. This comes from accepting $S_{IWW}$ as the appropriate geometric entropy, associated with the constant $v$ spatial slices of the light sheet, appearing on the RHS of Bousso bound, e.g., in \eqref{eq:finresult1}. The results in \cite{Wall:2015raa, Bhattacharyya:2021jhr} established that $S_{IWW}$ defines the entropy of a dynamical black hole away from stationarity, consistent with the linearized second law in higher derivative theories of gravity. The spatial density measure of how $S_{IWW}$ changes with $v$ defines for us the generalized expansion parameter of the null congruences on the horizon for higher derivative theories of gravity 
\begin{equation}
\partial_v S_{IWW}(v) =  \int  d^{D-2}x \, \sqrt{h(v,x)}~\vartheta_{IWW}(v,x) \, .
\end{equation}
Now, from \eqref{defIWW} it is obvious that 
\begin{equation}
\begin{split}
\vartheta_{IWW}(v,x) & = \frac{1}{\sqrt{h(v,x)}} \, \partial_v \left[\sqrt{h(v,x)}\left(1 +\alpha^2 {\cal J}_{IWW}(v,x)\right)\right] \\ 
& = \theta(v,x) + \alpha^2 {\cal Q}(v,x)\, , 
\end{split}
\end{equation}
where $\theta(v,x)$ from \eqref{eq:theta} defines the expansion parameter of the null congruences in Einstein's gravity. Note that the terms involving ${\cal J}^i$ in ${\cal Q}$ \eqref{eq:Evv} drop out in \eqref{defIWW} upon integrating on compact horizon slices. Therefore, it is quite naturally expected that for $v=0$ spatial slice $B$ of the light-sheet $L$, the assumption in \eqref{eq:assump2}, which was valid for general relativity, should be modified as 
 \begin{equation}\label{eq:assump2HD1} 
 s(0,x)\leq -\frac{\vartheta_{IWW}(0,x)}{4\hbar G_N}\,.
\end{equation}
which is same as \eqref{eq:assump2HD}.

%
%
%

\section{Discussions and future directions}\label{sec:conclude}

In this note, we have shown that the classical Bousso bound could possibly be violated if we add higher derivative corrections to Einstein's theory of gravity, even when the corrections are treated perturbatively. Then, we could modify the geometry side of the Bousso-bound so that these offending terms were taken care of. 

Let us summarize the framework under which our proof of the Bousso bound holds: \textit{First}, we adopt the effective field theory approach, where higher curvature terms are treated as small corrections (characterized by $\alpha^2$ ) to the two-derivative Einstein-Hilbert action. \textit{Second}, to illustrate the necessity of modifying the original Bousso bound, we consider spacetime metrics near the light sheet with slightly broken Killing symmetry, described by a perturbation parameter $\epsilon$. \textit{Third}, while the effective field theory parameter $\alpha^2$ and the perturbation parameter $\epsilon$ are independent, we focus on cases where the scale of the perturbation is set by the EFT parameter, specifically $\epsilon\sim\alpha^2$.

The modification we proposed is simply a replacement of the area term on the geometry side of the original Bousso bound by the Iyer-Wald-Wall entropy \cite{Wald:1993nt, Iyer:1994ys, Wall:2015raa, Bhattacharyya:2021jhr} integrated over the co-dimension two spatial slices of the light-sheet. Such a modification is well-motivated and could be anticipated from the recent progress on black hole thermodynamics establishing a linearized second law in the presence of higher derivative corrections. Recently, in \cite{Matsuda_2021}, similar modifications were guessed and subsequently justified, although either focussing on specific examples of higher derivative theories of gravity or being restricted to spherically symmetric configurations in Gauss-Bonnet theory. The key result in this note is to show that this guess indeed works out nicely and more generally to all diffeomorphism invariant gravitational theories beyond general relativity.

The crucial ingredient in establishing our result has been a particular off-shell structure of the equations of motion evaluated on the null light sheet. This structure is given in \eqref{eq:Evv} and was proved recently in \cite{Wall:2015raa, Bhattacharyya:2021jhr} for the null horizon of a slightly perturbed stationary black hole in arbitrary higher derivative theories of gravity. In these works, the off-shell structure of the equations of motion provided $S_{IWW}$, a definition of a dynamical black hole entropy consistent with the linearized second law. Intriguingly, the same off-shell structure of equations of motion modified the geometry side of the Bousso bound, i.e., the RHS of \eqref{eq:finresult1}. 

Our construction took care of those cases where even a perturbative higher derivative correction can violate the original Bousso bound - this can happen when the geometry around the light sheet has a slightly broken Killing symmetry (or a boost symmetry borrowing the terminology of \cite{Bhattacharyya:2021jhr}) as described in \S\,\ref{sec:violation}. Note that although the light sheet is, by definition, a null hypersurface, there is no a priori reason for it to be a Killing horizon. We are discussing light sheets with slightly broken Killing symmetry only because we could argue that for such geometric configurations, even the perturbative higher derivative corrections are threatening enough to invalidate the Bousso bound, and we have been successful in \emph{re-formulating} the bound that dealt with such situations.  

As we have described before, our proof (just like all other existing proofs of the classical Bousso bound) starts from physically motivated assumptions involving an inequality between the entropy and the energy flux (obtained in terms of the stress tensor) of the matter sector passing through the light-sheet. It is known that these assumptions and hence the classical proofs of Bousso bound works in a large class of situations \cite{Flanagan:1999jp, Bousso:2003kb}. However, counter-examples, where this bound can fail, were also given, focussing on situations where quantum effects become dominant \cite{Lowe:1999xk, Strominger:2003br}. Therefore, unless these assumptions are rigorously proved, these statements of the Bousso bound are just conjectures with a broad range of validity. 

Undoubtedly, it would be nice to justify these assumptions from first principles. Such attempts at a more rigorous formulation of the bounds have been made for two derivative general relativity. Identifying the shortcomings of the assumptions, they relied on theorems in quantum information theory to circumvent them. For example, the Bekenstein bound \cite{Bekenstein:1980jp} was rigorously formulated by Casini \cite{Casini_2008, Blanco:2013lea} based on certain inequalities satisfied by relative entropy in a generic relativistic quantum field theory. Similarly, for covariant entropy bounds associated with light-sheets (not necessarily with horizons of black holes), a quantum version of the Bousso bound was proved in \cite{Strominger:2003br} and later in \cite{Bousso:2014sda, Bousso:2014uxa}. They are similar in their approaches in considering the quantum entanglement entropy (or the von Neumann entropy) of matter fields present. However, they differ in the constructions of the proof and their applicability. It is interesting to check if their constructions can be extended to our setup, i.e., in situations where higher derivative corrections to general relativity are not negligible. One would need to generalize the concepts of von Neumann entropy of matter fields in the presence of higher derivative terms in gravity Lagrangian. We believe that the recent developments in the context of operator algebra in curved space-time \cite{Witten:2021jzq, Chandrasekaran:2022eqq, Kudler-Flam:2023qfl, Ali:2023bmm} would possibly be the correct approach to address these questions. However, at this moment, it is very ambitious and certainly beyond the scope of our paper. 

A related issue that naturally arises in the discussion of Bousso bound - specializing the light-sheet to the horizon of a black hole - is the generalized second law (GSL)\cite{Bekenstein:1974ax,Wall:2009wm,Sarkar:2013swa,Dhivakar:2023zqn}. The GSL states that the total black hole entropy and the von Neumann entropy of matter outside the horizon should never decrease. For general relativity coupled to semi-classical matter, it was argued in \cite{Flanagan:1999jp} that the classical Bousso bound readily implies the GSL under certain additional restrictions, such as the null energy condition being satisfied by the matter. Along these lines, in \cite{Matsuda_2021}, a GSL was proved as a by-product of the covariant entropy bound for spherically symmetric configurations in Einstein-Gauss-Bonnet gravity. Following the same chain of logic, our results in this paper will also imply a GSL for arbitrary higher derivative corrections to general relativity, at least up to a linear order in the non-Killing perturbations. However, as pointed out in \cite{Wall:2009wm}, these approaches to prove GSL implied by the Bousso bounds are restricted to situations where the entropy of the matter fields can be faithfully described by a local classical entropy current (termed as the hydrodynamic approximation) as adopted in the setup of proving a classical Bousso bound. Therefore, it is an interesting open direction to examine the validity of this hydrodynamic approximation in higher derivative theories of gravity to conclude a GSL in such theories. 


Following up on the issue of assigning a local (in the hydrodynamic sense) classical entropy current to matter fields in higher derivative theories of gravity, the question becomes even trickier if matter fields are non-minimally coupled to gravity. In two-derivative general relativity with minimal coupling, the matter stress tensor is well-defined as the variation of the matter Lagrangian with respect to the background metric. This stress tensor also satisfies various energy conditions, such as the null energy condition. However, the definition of the matter-Lagrangian and, therefore, the matter stress tensor becomes ambiguous in the presence of non-minimal coupling. With non-minimally coupled matter fields with gravity, the assumptions in \eqref{eq:assump1}, \eqref{eq:assump2}, which are crucial for a proof of the Bousso bound, can easily be violated. This is because these assumptions involving the entropy and energy flux of the matter imply null energy condition, which is well known to be violated with non-minimal coupling of matter with gravity \cite{Barcelo:2000zf, PhysRevD.54.6233, Chatterjee:2015uya, Wall:2018ydq}. 

For example, we could define the matter part of the Lagrangian as the part with at least one matter field. But then it can be shown that in the simplest classical situation, the null energy condition can be violated at the leading linearized order in the amplitude expansion about a stationary configuration with a null hyper-surface \cite{Biswas:2022grc}. Because of these unsettled puzzles, in this note, we restricted our proof to minimally coupled matter with all higher derivative corrections added only to the pure gravity Lagrangian. If we were to guess the modifications in the Bousso bound with non-minimally coupled matter field following the black hole thermodynamics \cite{Biswas:2022grc}, we would have absorbed some parts of the previously defined matter stress tensor (the parts that potentially violate the null energy condition) into the geometry side of the Bousso bound. This is definitely a possibility since, even with the non-minimal coupling, at the linearized order in the small breaking of the Killing symmetry of a light sheet, the equations of motion have the same desired off-shell structure \eqref{eq:Evv}. Following our analysis in this paper, this will imply a redefinition of the stress tensor by simply subtracting some parts. Therefore, in this way, even if we can prove the classical Bousso bound with non-minimal coupling, the immediate question that arises is the following. Does this redefined stress tensor satisfy all the properties that any stress tensor must fulfil, including its conservation? We shall leave this question as one of our future directions.

\section*{Acknowledgements}
We would like to thank Avipsa Chakrabarti, Prateksh Dhivakar, Bobby Ezhuthachan,  Krishna Jalan, Suman Kundu, Shuvayu Roy, and Aron Wall for useful discussions. PB would like to acknowledge the support provided by the grant CRG/2021/004539. NK acknowledges support from a MATRICS research grant (MTR/2022/000794) from the Science and Engineering Research Board (SERB), India. We would also like to acknowledge our debt to the people of India for their steady support to the research in the basic sciences.

\bibliographystyle{JHEP}
\bibliography{Bousso}

\end{document}